# Predicting the dynamic process and model parameters of the vector optical solitons in birefringent fibers via the modified PINN

**Gang-Zhou Wu** [1,#], **Yin Fang** [1,#], **Yue-Yue Wang** [1,*], **Guo-Cheng Wu** [2,*] **and Chao-Qing Dai** [1,*]

[1] *College of Optical, Mechanical and Electrical Engineering, Zhejiang A&F University, Lin'an 311300, China*

[2] *College of Math & Informat Science, Data Recovery Key Lab Sichuan Province, Neijiang Normal University, Neijiang 641100, China*

**Abstract.** A modified physics-informed neural network is used to predict the dynamics of optical pulses including one-soliton, two-soliton, and rogue wave based on the coupled nonlinear Schrödinger equation in birefringent fibers. At the same time, the elastic collision process of the mixed bright-dark soliton is predicted. Compared the predicted results with the exact solution, the modified physics-informed neural network method is proven to be effective to solve the coupled nonlinear Schrödinger equation. Moreover, the dispersion coefficients and nonlinearity coefficients of the coupled nonlinear Schrodinger equation can be learned by modified physics-informed neural network. This provides a reference for us to use deep learning methods to study the dynamic characteristics of solitons in optical fibers.

## 1. Introduction

The latest developments in the field of optical solitons have revealed the possibility of overcoming the distance limitation of solitary wave transmission systems. Since optical solitons are an ideal carrier to realize ultra-long distance and ultra-large capacity optical fiber communications, the study of the transmission characteristics of the optical solitons is regarded as a very valuable research hotspot in this field [1]. The nonlinear Schrödinger equation (NLSE) is used as a model to reveal the transmission characteristics of solitons in optical fibers. In order to better depict the specific phenomena in nonlinear optics, researchers introduced the coupled NLSE (CNLSE) to govern more complex behaviors in optical pulse transmission [2,3]. In recent years, the CNLSE has been widely used in many fields, including optical fiber communications, biophysics, and Bose-Einstein condensates [4]. The CNLSE is often used to describe the interaction between modes in optical fibers, for example, birefringent fibers [5] and two-mode fibers [6], and soliton wavelength division multiplexing, and multi-channel optical fiber networks, etc.

In the past years, the classical numerical and analytical methods were applied to investigate the soliton propagation characteristics in fibers. For the sake of the observation from different prediction results, numerical methods use more data and become more and more complex, which often leads to overly complex models and inevitably brings computational challenges. In recent years, with the continuous development of scientific research, how to better mine and further application of massive data in the field of science has made great progress. As one of the most important discoveries, neural networks and deep learning methods have been widely used in natural language processing [7], computer vision [8], image recognition and other fields [9]. Researchers believe that neural networks have great potential in model predictive control, data-driven prediction [10], multi physical / multi-scale modeling and simulation of physical processes [11]. Therefore, the neural network method is also widely used in the field of optics [12,13], including ultrafast nonlinear dynamics in optical fibers [14], modulation instability [15], and so on. At the same time, based on the universal approximation and high expressivity of neural networks, using neural networks to solve partial differential equations (PDEs) has become a hot field [16,17]. Neural networks have been used to solve Navier-Stokes equations [18], and Burgers equations [19], etc.

Recently, Raissi et al. proposed a physics-informed neural network (PINN) method to solve nonlinear PDEs [20]. It has been proved that it can accurately solve the forward problem of approximating the control mathematical model solution, and it can also accurately solve the inverse problem of inferring the model parameters from the observation data [20]. The physical equation is embedded into the neural network by the PINN, and the residual term is added to the loss function to constrain the admissible solution space. Through this transformation, the problem of solving nonlinear PDEs turns into the problem of optimizing loss function. The previous analytical methods have many formulas and complex calculation, and can't be applied to the prediction of model parameters. In contrast, neural network method has obvious advantages and potential.

Recently, Chen et al. applied the PINN method to study the propagation of solitons in water based on the KdV equation [21], and the propagation of solitons based on the traditional NLSE [22]. Yan et al. solved the forward and inverse problems of NLSE with the PT-symmetric harmonic potential [23] and also discussed the data-driven rogue wave solutions of defocusing NLSE [24]. However, the study of soliton propagation in birefringent fibers based on the CNLSE has not been reported by neural network methods. For solving nonlinear PDEs, some other neural network methods were gradually proposed [25,26]. We find that the training effect is bad by using a single network similar to that in Ref. [20] to solve coupled complex function partial differential equations, such as the CNLSE. To handle this problem, we must introduce a new network structure.

In this paper, a new network structure is proposed to study the data-driven solutions and

parameters discovery of CNLSE as [27]

$$iQ_{j,z} + Q_{j,tt} + 2(\sum_{l=1}^{2} \sigma_l |Q_l|^2)Q_j = 0, \qquad j = 1, 2, \qquad (1)$$

$Q_j$ is the slowly varying amplitude envelope of two interacting fiber modes, $z, t$ represents the normalized distance and delay time of the soliton propagating along the fiber in the birefringent fibers respectively, $\sigma_l$ is the sign of nonlinearity coefficients, including self-phase modulation and cross-phase modulation. For the sake of generality, this case is defined as

$$\sigma_l = \begin{cases} 1 & \text{for } l = 1, \\ \pm 1 & \text{for } l = 2. \end{cases} \qquad (2)$$

According to the different signs of nonlinearity coefficients, Eq. (1) can be divided into three cases. When $\sigma_1 = \sigma_2$, the CNLSE is the classical Manakov type. In this situation, the bright-bright soliton exists in the abnormal group velocity dispersion region in the fiber. Researchers found that bright soliton has an inelastic collision, including energy transfer and amplitude-related phase shift [30]. The Manakov model is very important in describing the process of pulse transmission in birefringent fibers. This model is a two-component vector extension of the NLSE. When $\sigma_1 = -\sigma_2$, the CNLSE (1) becomes the mixed one and its bright-dark soliton solution can be obtained. Researchers found that bright and dark solitons both have elastic collisions for the case of mixed two components [27, 28, 29]. In this situation, the phase shift of the bright soliton is affected by the dark soliton parameters.

The novelty of this article includes the following aspects. (i) The PINN method is improved so that it can be applied to the CNLSE; (ii) for the first time, the neural network method is used to predict the dynamic process of dark soliton; and (iii) in the model parameter prediction, the network is optimized from two aspects of the data sets and the proportion of the loss function, and the stability of the modified PINN is tested. These contents are our first attempts.

**2. Modified PINN method**

For different equations, the PINN needs to be specifically designed for different problems. The simple extension is not used for different problems. When solving the Navier–Stokes equation, Raissi used a network to find the solutions of two equations [20]. The network structure with two inputs in the input terminal and two outputs in the output terminal was used to represent two components respectively [20]. When the network used by Raissi for the Navier–Stokes equation in Ref. [20] is applied to solve coupled complex function partial differential equations, such as the CNLSE, we find that the training effect is bad because the number of neurons in a single network similar to that in Ref. [20] is not enough to characterize multiple complex solutions simultaneously. If the number of neurons in a single network is increased, the training time will substantially increase, which is something that we do not want to see.

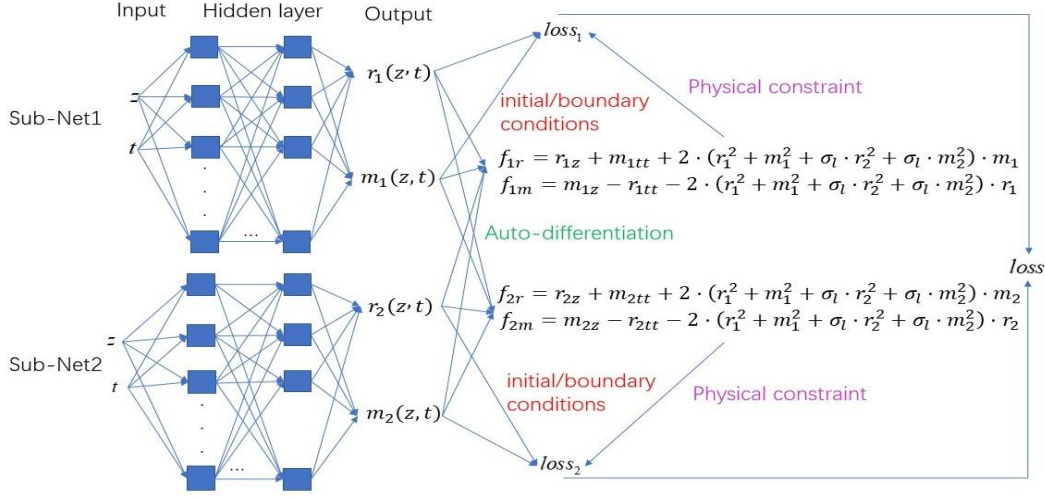

**Fig.1.** The framework of modified PINN for solving coupled equations.

In order to handle this problem, we modify the PINN method by improving the network structure, which includes two Sub-Nets. The network structure is shown in Fig.1. Each Sub-Net possesses two inputs and two outputs. They respectively correspond to the real part $r_{1,2}(z,t)$ and imaginary part $m_{1,2}(z,t)$ of each solution $Q_{1,2}$, which satisfies coupled complex function partial differential equations, such as the CNLSE, in the form

$$Q_{1z} + N(Q_1, Q_2, Q_{1t}, Q_{1tt}, \cdots) = 0,$$
$$Q_{2z} + N(Q_1, Q_2, Q_{2t}, Q_{2tt}, \cdots) = 0, \qquad (3)$$

where $N$ means arbitrary combination of linear and nonlinear terms of $Q_1, Q_2$. Neural networks share the same parameters (such as weights and biases), and are able to be trained by minimizing the mean squared error (MSE) loss. The MSE is generated by the initial boundary value conditions related to the feed forward neural network. For Eq. (3), after separating the real and imaginary parts, we can get

$$\begin{aligned}
f_{r1} &:= r_{1z} + N(r_1, r_2, m_1, m_2, m_{1t}, m_{1tt}, \cdots), \\
f_{r2} &:= r_{2z} + N(r_1, r_2, m_1, m_2, m_{2t}, m_{2tt}, \cdots), \\
f_{m1} &:= m_{1z} + N(r_1, r_2, m_1, m_2, r_{1t}, r_{1tt}, \cdots), \\
f_{m2} &:= m_{2z} + N(r_1, r_2, m_1, m_2, r_{2t}, r_{2tt}, \cdots).
\end{aligned} \qquad (4)$$

The shared parameters of the neural network are learned when the MSE loss

$$loss = MSE_0 + MSE_b + MSE_h, \qquad (5)$$

is minimized. Here

$$MSE_0 = \frac{1}{N_0} \sum_{i=1}^{N_0} (|r_1(z^i,t^i) - r_1^i|^2 + |m_1(z^i,t^i) - m_1^i|^2 + |r_2(z^i,t^i) - r_2^i|^2 + |m_2(z^i,t^i) - m_2^i|^2),$$

$$MSE_b = \frac{1}{N_b} \sum_{q=1}^{N_b} (|r_1(z^q,t^q) - r_1^q|^2 + |m_1(z^q,t^q) - m_1^q|^2 + |r_2(z^q,t^q) - r_2^q|^2 + |m_2(z^q,t^q) - m_2^q|^2), \qquad (6)$$

$$MSE_h = \frac{1}{N_f} \sum_{j=1}^{N_f} (|f_{r1}(z^j,t^j)|^2 + |f_{m1}(z^j,t^j)|^2 + |f_{r2}(z^j,t^j)|^2 + |f_{m2}(z^j,t^j)|^2),$$

where $\{r_1^i, r_2^i, m_1^i, m_2^i\}_{i=1}^{N_0}$ and $\{r_1^q, r_2^q, m_1^q, m_2^q\}_{q=1}^{N_b}$ represent the initial and boundary value data of $Q_1, Q_2$, $\{z^j, t^j\}_{j=1}^{N_f}$ is the collocation points on $f(z,t)$. In this paper, we choose $N_0 = 100, N_b = 100, N_f = 10000$.

There are various choices of the hidden layers and the number of neurons in the two Sub-Nets. In order to ensure that the two Sub-Nets spend similar time during training, we choose the following network structure: the hidden layers in Sub-Net1 have 6 layers with 40 neurons for each layer, and yet the hidden layers in Sub-Net2 have 7 layers with 30 neurons for each layer.

This is the first time to modify the PINN method to solve coupled complex function partial differential equations. We have improved the structure of the network by using two subnets. We find that this network structure has advantages in solving the CNLSE. Although the overall structure is more complicated than the single network in Ref. [20], but the operation time does not increases significantly.

## 3. Data-driven soliton solutions

We will discuss the CNLSE with Manakov type and mixed type, and predict dynamics of soliton solutions, including bright-bright one soliton solution, bright-bright soliton molecule solution, first-order and second-order rogue wave solutions, bright-dark one soliton solution, bright-dark soliton molecule solution, and interaction between bright-dark two solitons.

### 3.1. Solutions of Manakov-type CNLSE

We will use the improved PINN in Section 2 to predict the data-driven soliton solution of the Manakov-type CNLSE as follows [30,31]

$$Q_{1z} + Q_{1tt} + 2(|Q_1|^2 + |Q_2|^2)Q_1 = 0,$$
$$Q_{2z} + Q_{2tt} + 2(|Q_1|^2 + |Q_2|^2)Q_2 = 0.$$
(7)

### 3.1.1. Bright-bright one soliton solution

The exact bright-bright one soliton solution is as follows [31]

$$Q_1 = \frac{e^{t(0.2+0.99i)+z(1-0.1i)}}{1+0.5e^{0.4t+2z}}, Q_2 = \frac{e^{t(0.2+0.99i)+z(1-0.1i)}}{1+0.5e^{0.4t+2z}}, z \in [0,5], t \in [-6,6].$$
(8)

The sampling points set are obtained by means of pseudo-spectral method, and the exact bright-bright one-soliton solution is discretized into $[256 \times 201]$ data points. The initial and Dirichlet periodic boundary data are obtained by Latin hypercube sampling. The number of sampling points is given in Section 2.

Figure 2(a) shows the spatiotemporal dynamics of predicted bright-bright one soliton. The bright soliton exists in the anomalous group velocity dispersion region in the fiber, and the group velocity dispersion and phase modulation in the fiber are balanced, so that the bright soliton propagates stably in the fiber. In Fig. 2(b), we compare exact solution with the predicted solution of the bright-bright one soliton at $z=1.5, z=3.5$ for $Q_1(z,t)$ and $z=2, z=4.75$ for $Q_2(z,t)$, and we can see that they are completely coincident. The density diagram of error $Er_k = \hat{Q}_k(t,x) - Q_k(t,x)$ between the exact solution $Q_k(t,x)$ $(k=1,2)$ and the predicted solution $\hat{Q}_k(t,x)$ is shown in Fig. 2(c), where the large magnitude indicates the larger error. Fig. 2 (d) shows the curve of the loss function with the number of iterations. The loss function converges to $2.478 \times 10^{-5}$ after 8000 iterations. From the comparison of several aspects in Fig. 2, the prediction result is excellent and achieves our expectation.

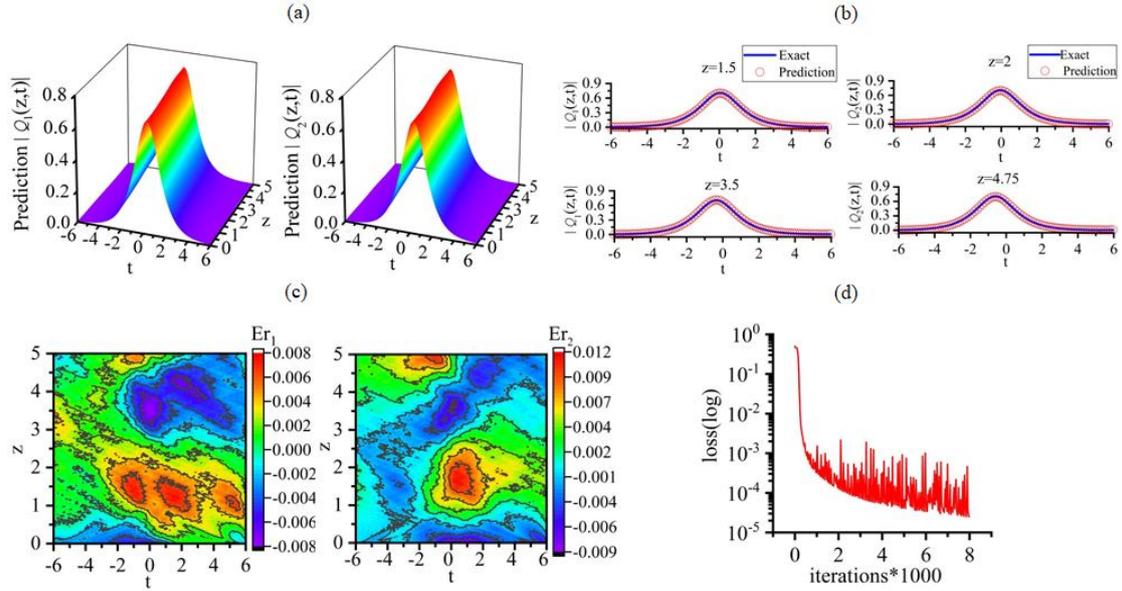

**Fig.2.** Bright-bright one soliton solutions $Q_1(z,t)$ and $Q_2(z,t)$ :(a) The predicted propagation process; (b) Comparisons between the exact (the blue solid line) and predicted results (the red circle) at different propagation distances; (c) The density diagram for the error between exact and predicted results; (d) Curve of the loss function v.s. the iterations.

Next, we discuss which adaptive activation function is an appropriate one when solving one soliton of the Manakov-type CNLSE. When the structure of the network remains unchanged and the number of learning is 6000, we study the influence of several conventional activation functions including sin, cos, tanh, swich functions on the convergence error and convergence speed of loss function. From Fig. 3, we find that the convergence of sin and switch functions is similar, the sin function is the best in the convergence rate, however the convergence effect of cos function is the worst.

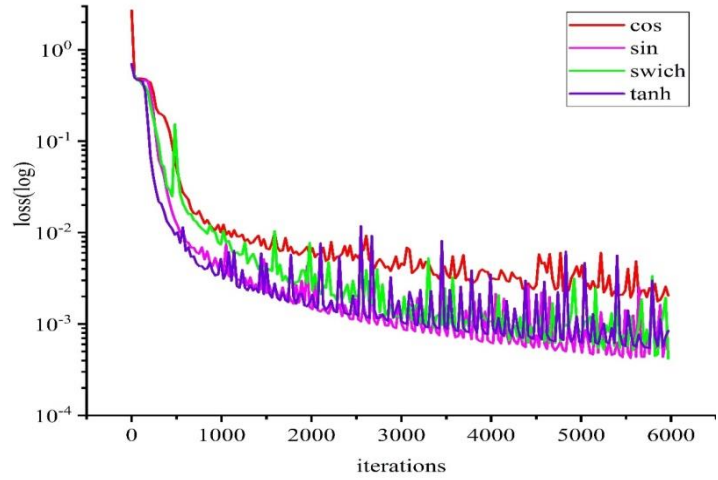

**Fig.3.** Convergence curves of loss function for different activation functions when solving one soliton of the Manakov-type CNLSE.

**Table.1.** Comparison on prediction of the one soliton solution for the Manakov-type CNLSE by using different activation functions.

| Activation functions \ Different parameters | time | $r_1$ | $m_1$ | $|Q_1|$ | $r_2$ | $m_2$ | $|Q_2|$ | loss |
|---|---|---|---|---|---|---|---|---|
| cos | 7507.69s | 63.43% | 38.66% | 14.9% | 5.42% | 3.81% | 2.87% | 0.00205 |
| sin | 6663.59s | 15.47% | 13.46% | 3.25% | 3.72% | 2.53% | 0.946% | 0.00041 |
| swich | 13320.4s | 23.10% | 19.03% | 4.84% | 3.90% | 2.49% | 1.26% | 0.00042 |
| tanh | 6617.60s | 22.24% | 18.28% | 4.66% | 3.65% | 2.58% | 1.03% | 0.00085 |

In Table 1, these percentages mean relative errors $(|\hat{N}-N|/N)\times 100\%$ with the predicted value $\hat{N}$ and exact value $N$, where $N$ includes $r_1, r_2, m_1, m_2, |Q_1|$ and $|Q_2|$. From Table 1, we find that sin, tanh and swich functions have little difference on the error of solution, cos function has the largest error, yet swich and sin functions have the best performance on the convergence of loss. However, due to the complexity of swich activation function, it needs longer time to find the derivative in back propagation. Therefore, we use the improved PINN with using sin function as the activation function in the following discussion.

### 3.1.2. Bright-bright two soliton molecule solution

The exact bright-bright two soliton molecule solution is as follows [30]

$$Q_1 = \frac{e^{1.1025it+1.05z+0.5} - 0.0053e^{it+3.1z+0.5} - 0.0052e^{1.1025it+3.05z+1.5} + 2e^{it+z+0.5}}{1+1.25e^{2z+1}+0.000034e^{4.1z+2}+0.4535e^{2.1z+1}+0.7139e^{2.05z-0.1025it+1}+0.7139e^{2.05z+0.1025it+1}},$$

$$Q_2 = \frac{e^{1.1025it+1.05z+0.5} + 0.0059e^{it+3.1z+0.5} + 0.0126e^{1.1025it+3.05z+1.5} + e^{it+z+0.5}}{1+1.25e^{2z+1}+0.000034e^{4.1z+2}+0.4535e^{2.1z+1}+0.7139e^{2.05z-0.1025it+1}+0.7139e^{2.05z+0.1025it+1}},$$

(9)

with the space-time region $z \in [0,2], t \in [-15,15]$.

Similarly to the prediction of bright-bright one soliton solution, the prediction result of bright-bright two-soliton molecule solution is also excellent and achieves the desired effects. In Fig. 4(a), the predicted bright-bright two soliton molecules maintain a stable propagation at a certain distance. The two solitons achieve the velocity resonance to form a bound state and keep an equal distance parallelly transmit without any interaction, and their amplitudes remain unchanged. In Fig. 4(b), we compare exact solution with the predicted solution of the bright-bright two soliton molecules at $z=0.8, z=1.9$ for $Q_1(z,t)$ and $z=0.6, z=1.4$ for $Q_2(z,t)$, and their shapes and structures fit very well. From the density pattern of error between exact solution and the predicted solution in Fig. 4(c), the error increases along the transmission distance. Fig. 4(d) shows the curve of the loss function of Eq. (9) and Eq. (10) with the increase of the iterations. The loss function converges to $6.437\times10^{-5}$ and $1.533\times10^{-4}$ after 12000 iterations, respectively.

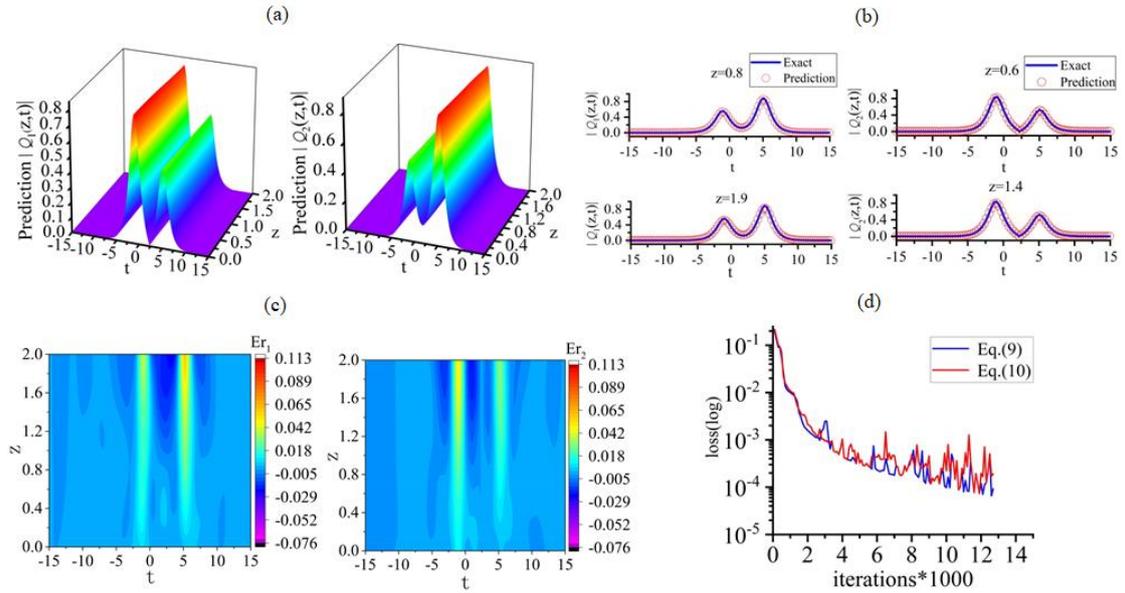

**Fig.4.** Bright-bright two soliton molecule solutions $Q_1(z,t)$ and $Q_2(z,t)$: (a) The predicted propagation process; (b) Comparisons between the exact (the blue solid line) and predicted results (the red circle) at different propagation distances; (c) The density pattern for the error between exact and predicted results; (d) Curve of the loss function v.s. the iterations.

Next, changing the coefficients of $z$ and $t$, the exact bright-bright two soliton molecule solution is as follows

$$Q_1 = \frac{2e^{1.44it+1.2z+1} - 0.0051e^{1.69it+3.7z+3} - 0.0052e^{1.44it+3.8z+3} + e^{1.69it+1.3z+1}}{1+0.86e^{2.4z+2}+0.000042e^{5z+4}+0.2959e^{2.6z+2}+0.48e^{2.5z-0.15it+2}+0.48e^{2.5z+0.15it+2}},$$

$$Q_2 = \frac{e^{1.44it+1.2z+1} + 0.0066e^{1.44it+3.8z+3} + 0.0147e^{1.69it+3.7z+3} + e^{1.69it+1.3z+1}}{1+0.86e^{2.4z+2}+0.000042e^{5z+4}+0.2959e^{2.6z+2}+0.48e^{2.5z-0.15it+2}+0.48e^{2.5z+0.15it+2}},$$

(10)

with the space-time region $z\in[0,2], t\in[-15,15]$.

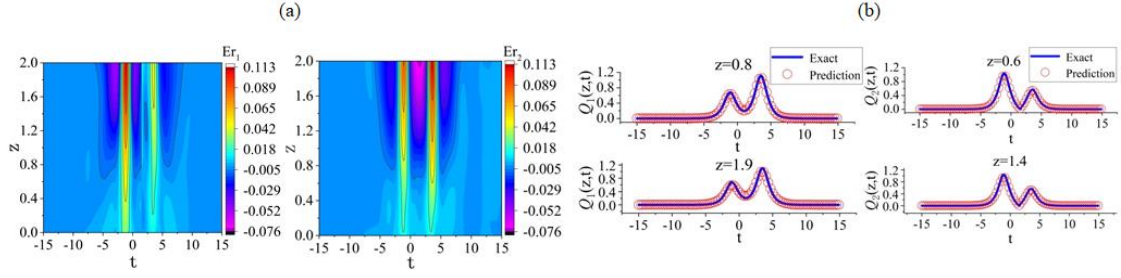

**Fig.5.** Bright-bright two soliton molecule solutions $Q_1(z,t)$ and $Q_2(z,t)$ : (a) The density pattern for the error between exact and predicted results; (b) Comparisons between the exact (the blue solid line) and predicted results (the red circle) at different propagation distances.

In Fig.5, changing the value of soliton parameters (coefficients of $z$ and $t$ in the exponential) to a new solution as a data set. Fig. 5 (a) shows the density plots of the absolute errors. In Fig. 5 (b), we compare the exact solution and predicted solution of the bright-bright two soliton molecules at $z=0.8, z=1.9$ for $Q_1(z,t)$ and $z=0.6, z=1.4$ for $Q_2(z,t)$. Compared with Fig. 4, the error increases, but the law of the increase of the error along the transmission distance does not change. Therefore, changing the coefficients of $z$ and $t$ in the exponential cannot bring any favourable impact on the results.

### 3.1.3. First-order rogue wave

The exact first-order rogue wave solution is as follows [32]

$$Q_1 = 0.6e^{1.44it}(1 - \frac{4+11.52it}{2.88(0.0116t-z)^2 + 8.2944t^2 + 1}),$$
$$Q_2 = 0.6e^{1.44it}(1 - \frac{4+11.52it}{2.88(0.0116t-z)^2 + 8.2944t^2 + 1}),$$
(11)

with the space-time region $z \in [-1,1], t \in [-4,4]$.

Similarly to the prediction of soliton solution above, the prediction result of first-order rogue wave is also excellent and achieves the desired effects. In Fig. 6(a), the predicted first-order rogue wave exhibits high amplitude in the optical fiber. Due to the modulation instability in the optical fiber, the weak modulation on the plane wave can produce exponential growth along the transmission distance, and result in rogue waves. In Fig. 6(b), we compare exact solution with the predicted solution of the first-order rogue wave at $z=-0.4, z=0.4$ for $Q_1(z,t)$ and $z=-0.6, z=0.8$ for $Q_2(z,t)$, and they have a good match. From the 3D diagram of error between exact solution and the predicted solution in Fig. 6(c), the error on both sides of the boundary is relatively large. Fig. 6(d) exhibits the curve of the loss function with the increase of the iterations. The loss function converges to $1.535 \times 10^{-3}$ after 8000 iterations.

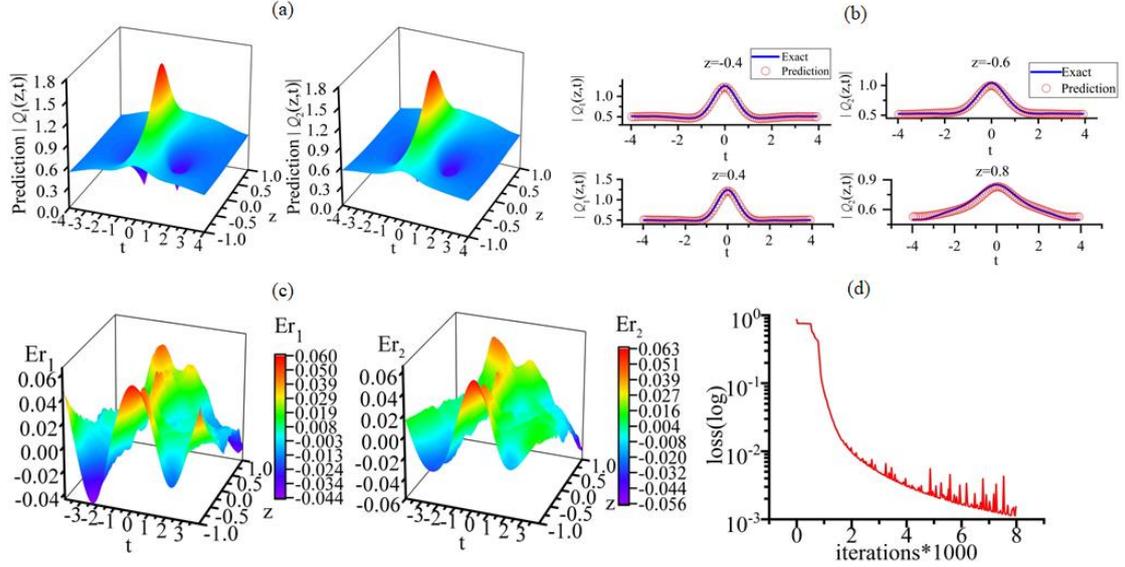

**Fig.6.** First-order rogue wave solutions $Q_1(z,t)$ and $Q_2(z,t)$ : (a) The predicted propagation process; (b) Comparisons between the exact (the blue solid line) and predicted results (the red circle) at different propagation distances; (c) The 3D diagram for the error between exact and predicted results; (d) Curve of the loss function v.s. the iterations.

### 3.1.4. Second-order rogue wave

The exact second-order rogue wave solution is as follows [33]

$$Q_1 = 0.7071e^{2it}(1-\frac{12t^2(4z^2+3)+it(16t^2(4z^2+1)+128t^4-12z^2+8z^4-5)+160t^4+3z^2+2z^4-0.5}{0.0104z^4(768t^2+48)+0.125z^2(16t^2-3)^2+16.5t^2+72t^4+42.6667z^6+0.0938}), \quad (12)$$

$$Q_2 = 0.7071e^{2it}(1-\frac{12t^2(4z^2+3)+it(16t^2(4z^2+1)+128t^4-12z^2+8z^4-5)+160t^4+3z^2+2z^4-0.5}{0.0104z^4(768t^2+48)+0.125z^2(16t^2-3)^2+16.5t^2+72t^4+42.6667z^6+0.0938},$$

with the space-time region $z \in [-1.5, 1.5], t \in [-4, 4]$, and the prediction result is also excellent in Fig.7.

Fig. 7(a) shows the predicted dynamical characteristics of the second-order rogue wave. The structural feature of the aggregated second-order rogue waves is a single peak structure, which can be regarded as the nonlinear superposition of three first-order rogue waves at the maximum peak. In Fig. 7(b), we compare exact solution with the predicted solution of the second-order at $z = -0.6, z = 0.6$ for $Q_1(z,t)$ and $z = -0.3, z = 0.3$ for $Q_2(z,t)$, and they fit very well. From the density pattern of error between exact solution and the predicted solution in Fig. 7(c), the error on both sides of the boundary is relatively large. Fig. 7(d) displays the curve of the loss function v.s. the iterations. The loss function converges to $4.429 \times 10^{-3}$ after 20000 iterations.

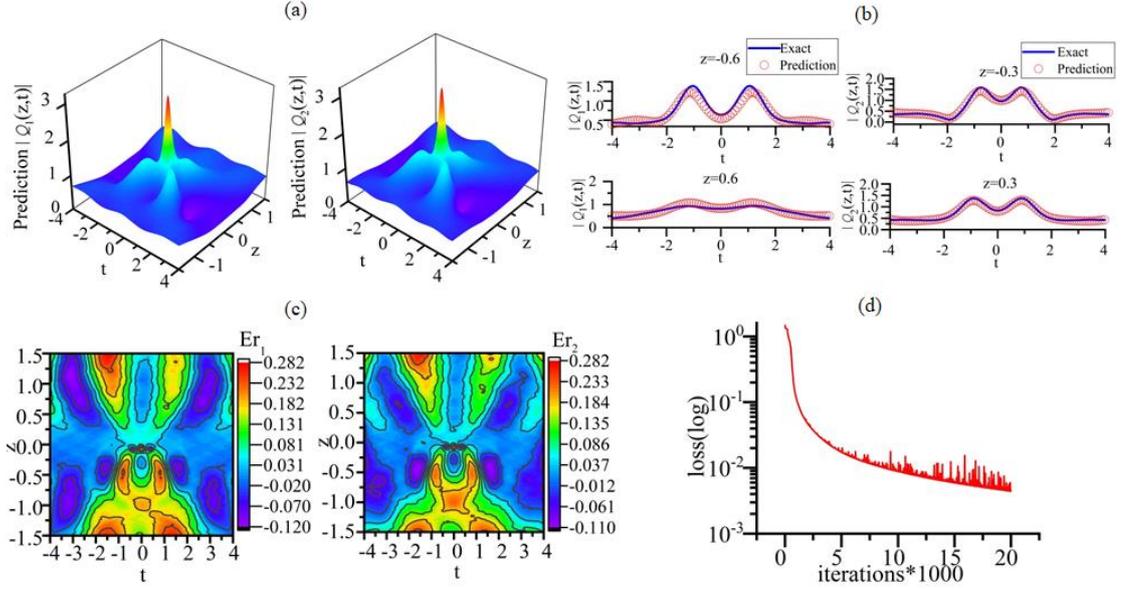

**Fig.7.** Second-order rogue wave solutions $Q_1(z,t)$ and $Q_2(z,t)$: (a) The predicted propagation process; (b) Comparisons between the exact (the blue solid line) and predicted results (the red circle) at different propagation distances; (c) The density diagram for the error between exact and predicted results; (d) Curve of the loss function v.s. the iterations.

### 3.2. Solutions of Mixed-type CNLSE

We will use the improved PINN in Section 2 to predict the data-driven soliton solution of the mixed-type CNLSE as follows [27, 28, 29]

$$\begin{aligned} Q_{1z} + Q_{1tt} + 2(|Q_1|^2 - |Q_2|^2)Q_1 &= 0, \\ Q_{2z} + Q_{2tt} + 2(|Q_1|^2 - |Q_2|^2)Q_2 &= 0. \end{aligned} \quad (13)$$

#### 3.2.1. Bright-dark one soliton solution

The exact bright-dark one soliton is as follows [27]

$$Q_1 = \frac{15e^{3.02it+2z}}{4.0009e^{4z}+4},$$
$$Q_2 = \frac{7e^{0.2iz-1.02it}(1-e^{4z}(0.9804-0.1981i))}{10.0022e^{4z}+10}, z \in [0,1], t \in [-8,8]. \quad (14)$$

In Fig. 8(a), the predicted spatiotemporal dynamics of bright-dark one soliton is displayed. The bright-dark solitons are trained, and they transmit stably. In Fig. 8(b), we compare exact solution with the predicted solution of the bright-dark one soliton at $z=0.3, z=0.7$ for $Q_1(z,t)$ and $z=0.4, z=0.95$ for $Q_2(z,t)$, and they are completely coincident. The 3D diagram of error between exact solution and the predicted solution is portrayed in Fig. 8(c), where the error increases along the transmission distance. Fig. 8(d) exhibits the curve of the loss function v.s. the iterations, where the loss function converges to $4.307 \times 10^{-4}$ after 20000 iterations. From the comparison of several aspects in Fig. 7, the prediction result is good and achieves our expected effect.

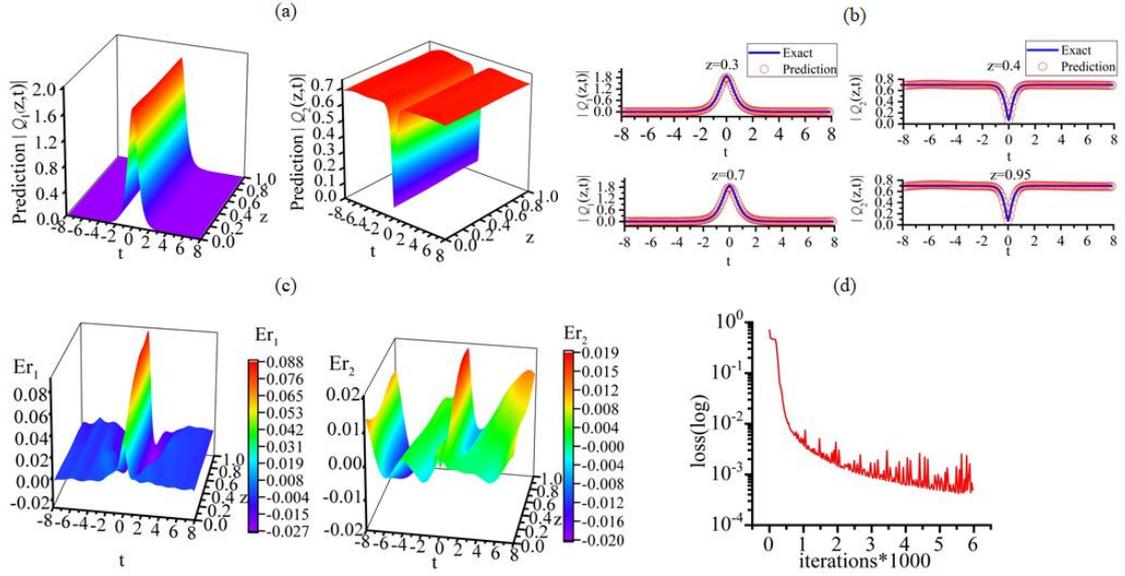

**Fig.8.** Bright-dark one soliton solutions $Q_1(z,t)$ and $Q_2(z,t)$: (a) The predicted propagation process; (b) Comparisons between the exact (the blue solid line) and predicted results (the red circle) at different propagation distances; (c) The 3D diagram for the error between exact and predicted results; (d) Curve of the loss function v.s. the iterations.

### 3.2.2. Bright-dark two soliton molecules solution

The exact bright-dark two soliton molecule solution is as follows [27]

$$Q_1 = \frac{e^{0.5it+1.4z}(37.1681+1.9318i) + e^{0.94it+1.2z} + e^{0.94it-0.8z}(81.9-5.5941i) + 1.4e^{0.5it-z}}{1 + e^{0.2z-0.44it}(29.345+1.7647i) + e^{0.44it+0.2z}(29.345-1.7647i) + 0.6451e^{-2z} + 0.2089e^{2.4z} + 2246.9e^{0.4z}},$$

$$Q_2 = \frac{e^{0.2iz-0.54it}(1 + e^{0.2z-0.44it}(24.5387+2.2716i) + e^{0.44it+0.2z}(35.0561-0.9754) - e^{-2z}(0.5954+0.2481i) - e^{2.4z}(0.1976-0.0678) + e^{0.4z}(2242.218+144.8099i))}{2 + e^{0.2z-0.44it}(58.6900+3.5294i) + e^{0.44it+0.2z}(58.69-3.5294i) + 1.29e^{-2z} + 0.4178e^{2.4z} + 4493.7e^{0.4z}},$$

(15)

with the space-time region $z \in [0,1.6], t \in [-17.5,17.5]$, and the prediction result of this solution is also very good in Fig. 9.

Fig. 9(a) shows the predicted dynamic characteristics of bright-dark two soliton molecules. The propagation velocity of the solitons in the two components is the same, and two solitons form bound state solitons (also called soliton molecule). The amplitude of two solitons does not change significantly, and they keep equal interval along the propagation distance. In Fig. 9(b), we compare exact solution with the predicted solution of the bright-dark one soliton at $z=0.48, z=1.12$ for $Q_1(z,t)$ and $z=0.64, z=1.44$ for $Q_2(z,t)$, and they get a consistent match. Fig. 9(c) shows the 3D diagram of error between exact solution and the predicted solution, and the error increases along the transmission distance. Fig. 9(d) shows the curve of the loss function with the add of the iterations. The loss function converges to $9.649 \times 10^{-5}$ after 12000 iterations.

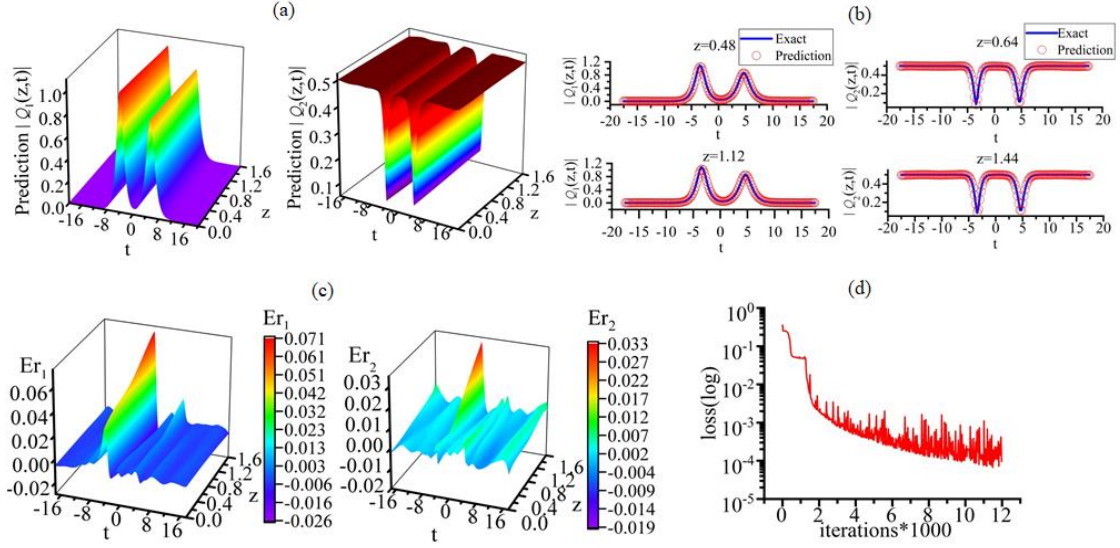

**Fig.9.** Bright-dark two soliton molecules solutions $Q_1(z,t)$ and $Q_2(z,t)$ :(a) The predicted propagation process; (b) Comparisons between the exact (the blue solid line) and predicted results (the red circle) at different propagation distances; (c) The 3D diagram for the error between exact and predicted results; (d) Curve of the loss function v.s. the iterations.

### 3.2.3. Bright-dark two soliton interaction

The exact bright-dark two soliton interaction solution is as follows [27]

$$Q_1 = \frac{1.4e^{-t(2.2+1.19i)-z(1+1.1i)} + e^{z(1.4-1.1)-t(11.8+1.19i)}(0.0254-0.3389i) + e^{z(1.2+2i)-t(4.8+3.54i)} + e^{-t(9.2+3.54i)-z(0.8-2i)}(0.2561-0.7040i)}{1+e^{0.4z-14t}(0.1890+0.3250\times10^{-17}i) - e^{z(0.2+3.1i)-t(7+2.35i)}(0.1263+0.0174i) + -e^{z(0.2-3.1i)-t(7-2.35i)}(0.1263-0.0174i) + 0.1939e^{2.4z-9.6t} + 0.5991e^{-2z-4.4t}},$$

$$Q_2 = \frac{7e^{0.2iz-1.02it}(1+e^{2.4z-9.6t}(0.0746-0.1790i) - e^{0.4z-14t}(0.15+0.115) + e^{z(0.2+3.1i)-t(7+2.35i)}(0.0753-0.1504i) + e^{z(0.2-3.1i)-t(7-2.35i)}(0.0183-0.0949) + e^{-2z-4.4t}(0.1537+0.5791i))}{10+1.9391e^{2.4z-9.6t} + e^{0.4z-14t}(0.1890+0.3250\times10^{-17}i) - e^{z(0.2+3.1i)-t(7-2.35i)}(1.2631+0.1742i) - e^{z(0.2-3.1i)-t(7-2.35i)}(1.2631-0.1742i) + 6e^{-4.4t-2z}},$$

(16)

with the space-time region $z \in [-1,1], t \in [-15.4, 15.4]$, and the prediction result is excellent and achieves the desired expectation in Fig. 10.

In Fig. 10 (a), the predicted dynamic characteristics of bright-dark two soliton interaction are displayed. The predicted results are consistent with the properties of exact soution. In the case of two components, there are elastic collisions between bright solitons and dark solitons, that is, they still maintain the original direction and amplitude after interaction, and there is no energy transfer. In Fig. 10(b), we compare exact solution with the predicted solution of the bright-dark two soliton interaction at $z = -0.4, z = 0.4$ for $Q_1(z,t)$ and $z = -0.2, z = 0.9$ for $Q_2(z,t)$, and they are in good agreement. Fig. 10(c) exhibits the density diagram of error between exact solution and the predicted solution. Fig. 10(d) shows the curve of the loss function with the add of the iterations. The loss function converges to $6.192 \times 10^{-4}$ after 12000 iterations.

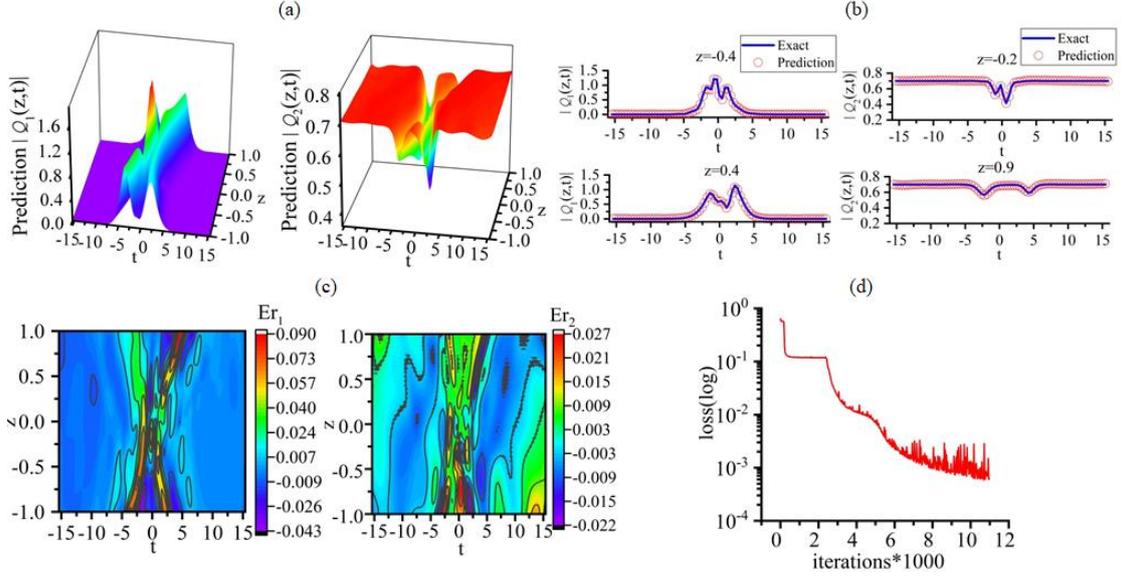

**Fig.10.** Bright-dark two soliton interaction solutions $Q_1(z,t)$ and $Q_2(z,t)$ : (a) The predicted propagation process; (b) Comparisons between the exact (the blue solid line) and predicted results (the red circle) at different propagation distances; (c) The density pattern for the error between exact and predicted results; (d) Curve of the loss function v.s. the iterations.

## 4. Parameter prediction of physical model

In this section, we will consider the parameter discovery problem of a data-driven CNLSE model. The Manakov-type equation is as follows

$$iQ_{1z} + \lambda_1 Q_{1tt} + \lambda_2 (|Q_1|^2 + |Q_2|^2)Q_1 = 0, \\ iQ_{2z} + \lambda_3 Q_{2tt} + \lambda_4 (|Q_1|^2 + |Q_2|^2)Q_2 = 0, \tag{17}$$

where slowly varying envelopes $Q_1, Q_2$ contain real parts $r_1, r_2$ and imaginary parts $m_1, m_2$, and variables $\lambda_1, \lambda_2, \lambda_3, \lambda_4$ are the unknown dispersion coefficients and nonlinearity coefficients to be trained.

The physical models $f_{1,2}(z,t)$ for training are defined as

$$f_1 := iQ_{1z} + \lambda_1 Q_{1tt} + \lambda_2 (|Q_1|^2 + |Q_2|^2)Q_1, \\ f_2 := iQ_{2z} + \lambda_3 Q_{2tt} + \lambda_4 (|Q_1|^2 + |Q_2|^2)Q_2, \tag{18}$$

which contain real and imaginary parts as $f_{1,2}(z,t) = f_{r1,2}(z,t) + if_{m1,2}(z,t)$, and we get

$$f_{r1} := r_{1z} + \lambda_1 m_{1tt} + \lambda_2 (r_1^2 + m_1^2 + r_2^2 + m_2^2)m_1, \\ f_{m1} := m_{1z} - \lambda_1 r_{1tt} - \lambda_2 (r_1^2 + m_1^2 + r_2^2 + m_2^2)r_1, \\ f_{r2} := r_{2z} + \lambda_3 m_{2tt} + \lambda_4 (r_1^2 + m_1^2 + r_2^2 + m_2^2)m_2, \\ f_{m2} := m_{2z} - \lambda_3 r_{2tt} - \lambda_4 (r_1^2 + m_1^2 + r_2^2 + m_2^2)r_2. \tag{19}$$

Our purpose is to minimize the MSE to obtain the approximate value of these unknown coefficients via the PINN method. The loss function of the neural network is defined as

$$loss = W_{rm} MSE_{rm} + W_f MSE_f, \tag{20}$$

where $W_{rm}, W_f$ are the weights of the initial/boundary and residual terms, and $MSE$ represents the mean square error

$$MSE_{rm} = \frac{1}{N_s} \sum_{\Re=1}^{N_s} (|r_1(z^\Re, t^\Re) - r_1^\Re|^2 + |m_1(z^\Re, t^\Re) - m_1^\Re|^2 + |r_2(z^\Re, t^\Re) - r_2^\Re|^2 + |m_2(z^\Re, t^\Re) - m_2^\Re|^2),$$

$$MSE_f = \frac{1}{N_s} \sum_{\Re=1}^{N_s} (|f_{r1}(z^\Re, t^\Re)|^2 + |f_{m1}(z^\Re, t^\Re)|^2 + |f_{r2}(z^\Re, t^\Re)|^2 + |f_{m2}(z^\Re, t^\Re)|^2).$$
(21)

Using the pseudo spectral method, the one soliton solution of the Manakov-type CNLSE is discretized into a data set with the space-time region $(z,t) \in [-6,6] \times [0,5]$, and thus the number of points in the data set is $[256 \times 201]$. The correct coefficients of the Manakov-type CNLSE are $\lambda_1 = 1, \lambda_2 = 2, \lambda_3 = 1, \lambda_4 = 2$. For the sake of the train for unknown dispersion coefficients and nonlinearity coefficients, we use the 6-layer deep neural network and 50 neurons for each layer with the number of sampling points $N_s = 5000$. Table 2 shows the training results of these unknown coefficients and the relative error $R_e = (|\hat{\lambda}_\tau - \lambda_\tau|/\lambda_\tau) \times 100\%$ ($\tau$=1,2,3,4) with the predicted value $\hat{\lambda}_\tau$ and exact value $\lambda_\tau$. From it, we think that the training effect is quite excellent.

**Table.2.** Comparison of correct model and identified model obtained by the PINN

| Item<br>CNLSE | Model | Relative error $R_e$ | | | |
|---|---|---|---|---|---|
| | | $\lambda_1$ | $\lambda_2$ | $\lambda_3$ | $\lambda_4$ |
| Correct equations | $iQ_{1z} + Q_{1tt} + 2(|Q_1|^2 + |Q_2|^2)Q_1 = 0$<br>$iQ_{2z} + Q_{2tt} + 2(|Q_1|^2 + |Q_2|^2)Q_2 = 0$ | / | / | / | / |
| Identified equations | $iQ_{1z} + 1.000056 Q_{1tt} + 2.000033(|Q_1|^2 + |Q_2|^2)Q_1 = 0$<br>$iQ_{2z} + 1.000032 Q_{2tt} + 2.000016(|Q_1|^2 + |Q_2|^2)Q_2 = 0$ | 0.005591% | 0.001657% | 0.003195% | 0.000799% |

Next we will analyze the training effect of the neural network from different perspectives, such as the category of data sets, the weight of loss function and the anti-interference ability.

We analyze the changes of the unknown coefficients and loss function with the iterations when we use different solutions such as one soliton, two soliton and rogue wave solutions as data sets in the inverse problem. Fig. 11(a)-(d) indicates the variation of the unknown coefficients with the iterations under different data sets. We find that the volatility of the unknown coefficient is the smallest when a one soliton is used as the data set, and yet this volatility is the largest when the rogue wave is used as the data set. Fig. 11(e) portrays the change of the loss function for different data sets with the increase of the number of iterations. It implies that the convergence speed of the loss function is fastest and the convergence error is smallest when a one soliton is used as the data set, and yet these performances are the worst when the rogue wave is used as the data set. From these results, we find that when predicting the physical parameters of the model, the simpler the structure of the data set, the better the training effect. Therefore, we choose the one soliton solution as the data set for training the inverse problem.

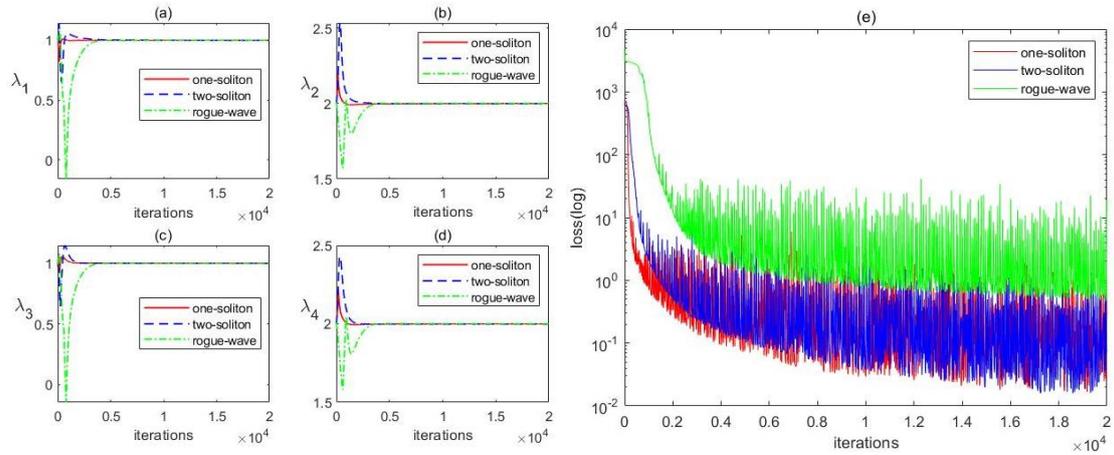

**Fig.11.** (a)-(d) the variation of unknown coefficients $\lambda_1, \lambda_2, \lambda_3, \lambda_4$ and (e) the variation of loss function with the iterations for different data sets.

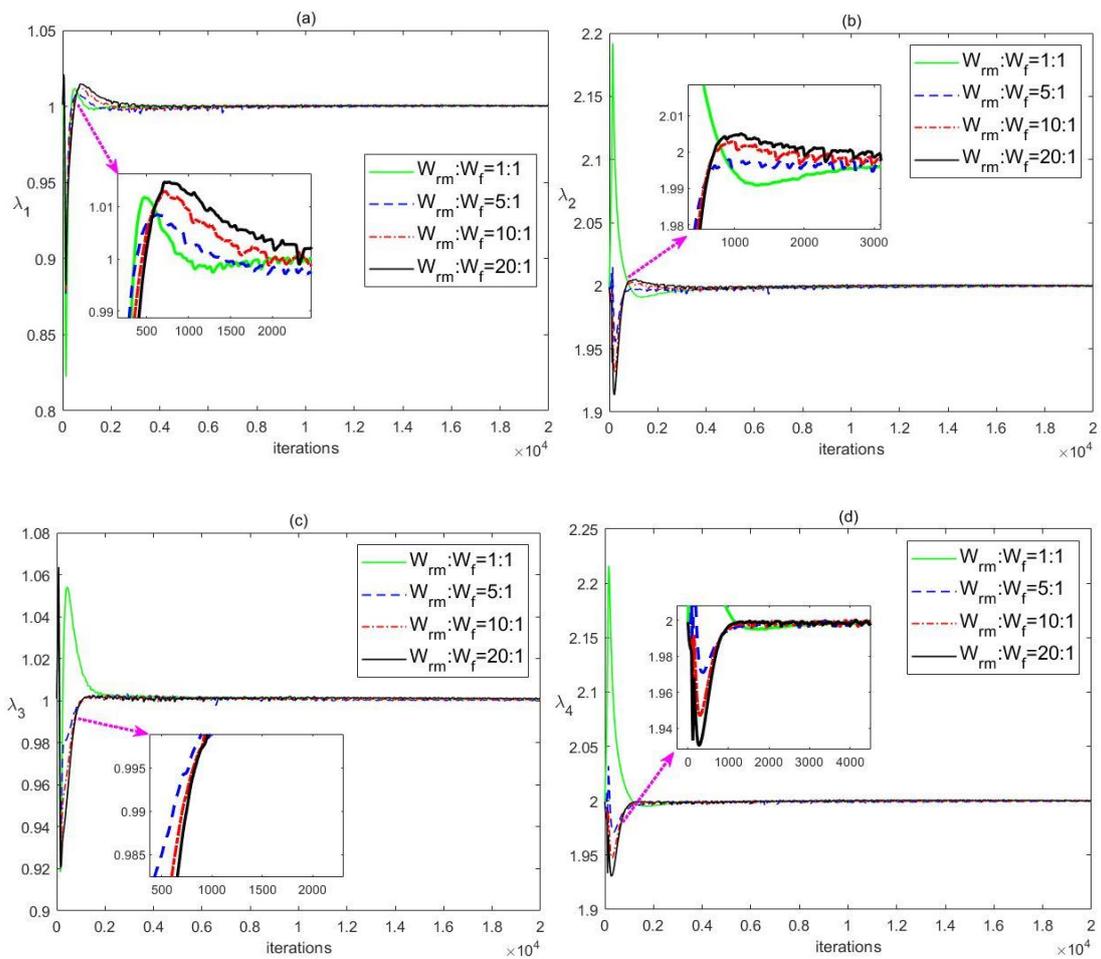

**Fig.12.** Variation of unknown coefficients $\lambda_1, \lambda_2, \lambda_3, \lambda_4$ with the iterations for different weights of the loss function.

In order to find the appropriate component weights of the model and significantly improve the prediction accuracy, we analyze the changes of unknown parameters with the number of iterations by changing the weights of the initial/boundary and residual terms of the loss function. When using the gradient descent method to train the PINN, by adjusting the weight of different components in the loss function, the gradient values of different components in the loss function tend to be consistent, which can effectively improve the convergence effect. In Fig. 12, for the same number of iterations, when $W_{rm}:W_f = 5:1$, the fluctuation of all unknown coefficients is the smallest and the convergence is faster. Therefore, for this model, $W_{rm}:W_f = 5:1$ is the best choice.

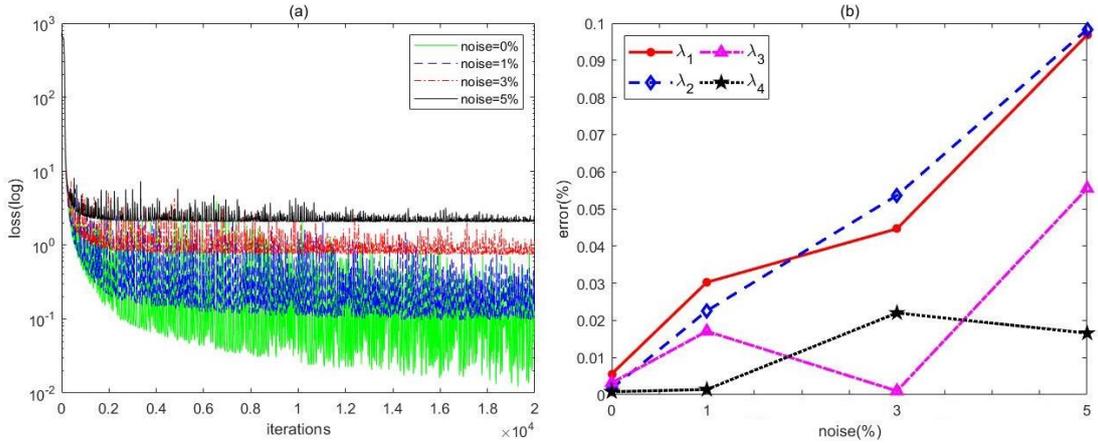

**Fig.13.** (a) The change of the loss function with the iterations and (b) the change of the error for the unknown coefficients $\lambda_1, \lambda_2, \lambda_3, \lambda_4$ for different interference noises.

For the sake of the test of the stability for the neural network, we add the interference noise to the sampling points. Here the 0%-5% noises are added to the sampling points respectively. In Fig. 13(a), with the increase of the noise, the convergence speed of the loss function significantly slows down, and the convergence error significantly increase. Fig. 13(b) shows the training error of unknown coefficients for different interference noises. We find that the improved PINN can accurately predict the unknown coefficients, and the error is still within an acceptable range even if the sampled data is corrupted by 5% noise. This shows that the neural network we proposed in Section 2 has a good stability.

## 5. Conclusion

In conclusion, we propose an improved PINN structure and analyze the activation function, and thus give a new network structure suitable for the coupled complex function PDEs. Using the modified PINN method, we predict the soliton transmission process in the frame of the CNLSE, including bright-bright one soliton, bright-bright two soliton molecules, first-order and second-order rogue waves, bright-dark one soliton, bright-dark two soliton molecule, and bright-dark two soliton interactions. To the best of our knowledge, it is the first report to predict the transmission process of dark solitons by using the improved PINN, and find the dynamic characteristics of dark solitons predicted by neural network. We find that the errors of bright-bright one soliton and soliton molecule, bright-dark one soliton and soliton molecule increase along the transmission distance. The errors at the beginning and end of the first-order and second-order rogue waves are obviously larger than those in the middle, while the error for the interaction of bright-dark two soliton has no obvious rule.

For the inverse problem, we mainly conduct research on the model parameter prediction from two aspects: the data set and the weight of the loss function. For different data sets, it can be seen that the training effects of neural networks have obvious differences. For the CNLSE, one soliton solution is an optimal choice to construct the data set. This also conforms to the idea that the progress of machine learning has been driven by improving the performance of data sets. Moreover, we find a suitable loss function weight, using which can significantly improve the convergence speed of the neural network. We have also studied the capacity of resisting interference for the proposed neural network structure. Our prediction results are still excellent against high noise interference, which indicates the good stability of our proposed neural network.

Although we have obtained excellent results, there are still issues worthwhile to study. For such, how to optimize the neural network? How to combine our results with experiments? Whether this modified PINN method is applicable to other models? These issues will be further studied in our future works.